\documentclass[11pt]{article}

\usepackage[preprint]{acl}
\usepackage{times}
\usepackage{latexsym}
\usepackage[T1]{fontenc}
\usepackage[utf8]{inputenc}
\usepackage{microtype}
\usepackage{inconsolata}
\usepackage{graphicx}
\usepackage{xspace}
\usepackage{booktabs}
\usepackage{multirow}
\usepackage{makecell}
\usepackage[table]{xcolor}

\renewcommand{\arraystretch}{1.15}
\newcommand{\approach}{{CrossSWE}\xspace}

\title{\approach{}: Benchmarking Cross-Repository Joint Development for Coding Agents}
\author{
  Silin Chen\textsuperscript{1}\begin{NoHyper}\thanks{Equal contribution.}\end{NoHyper}\quad
  Yuling Shi\textsuperscript{1*}\quad
  Heng Lian\textsuperscript{2}\quad
  Zihan Zhang\textsuperscript{1} \\
  \bfseries Xiaodong Gu\textsuperscript{1}\quad
  Haotian Lin\textsuperscript{1}\quad
  Zhensu Sun\textsuperscript{3}\begin{NoHyper}\thanks{Corresponding author.}\end{NoHyper} \\[2pt]
  \normalfont\normalsize\textsuperscript{1}\,Shanghai Jiao Tong University\qquad
  \textsuperscript{2}\,Xidian University \\
  \normalfont\normalsize\textsuperscript{3}\,Singapore Management University \\[2pt]
  \normalfont\small\texttt{\{yuling.shi, sjtu2023zzh, xiaodong.gu, hyperlynnx\}@sjtu.edu.cn} \\
  \normalfont\small\texttt{cslsolow@gmail.com\quad lianheng23@163.com\quad zssun@smu.edu.sg}
}

\begin{document}
\maketitle

\begin{abstract}
As coding agents achieve increasingly high success rates on software engineering benchmarks, recent efforts have sought to increase task complexity through longer horizons and more extensive code modifications. However, real-world software development often spans multiple interdependent repositories, introducing additional challenges in dependency reasoning, change propagation, and interface consistency. Rather than further increasing complexity within individual repositories, we introduce \approach{}, a benchmark for cross-repository joint development comprising 171 real-world instances spanning 165 repositories, 20 technical domains, and 13 programming languages. \approach{} verifies the necessity of coordinated changes by testing whether partial repository modifications are insufficient to achieve the intended behavior. Our evaluation of five LLMs reveals substantial challenges, with the strongest model, GPT-6.1 Sol, achieving only 58.48\% Pass@1. Furthermore, cross-repository inconsistencies emerge as the most prevalent failure mode (44.2\%), highlighting the difficulty of maintaining consistency across interdependent repositories.
\end{abstract}


\begin{figure}[t]
  \centering
  \includegraphics[width=\columnwidth]{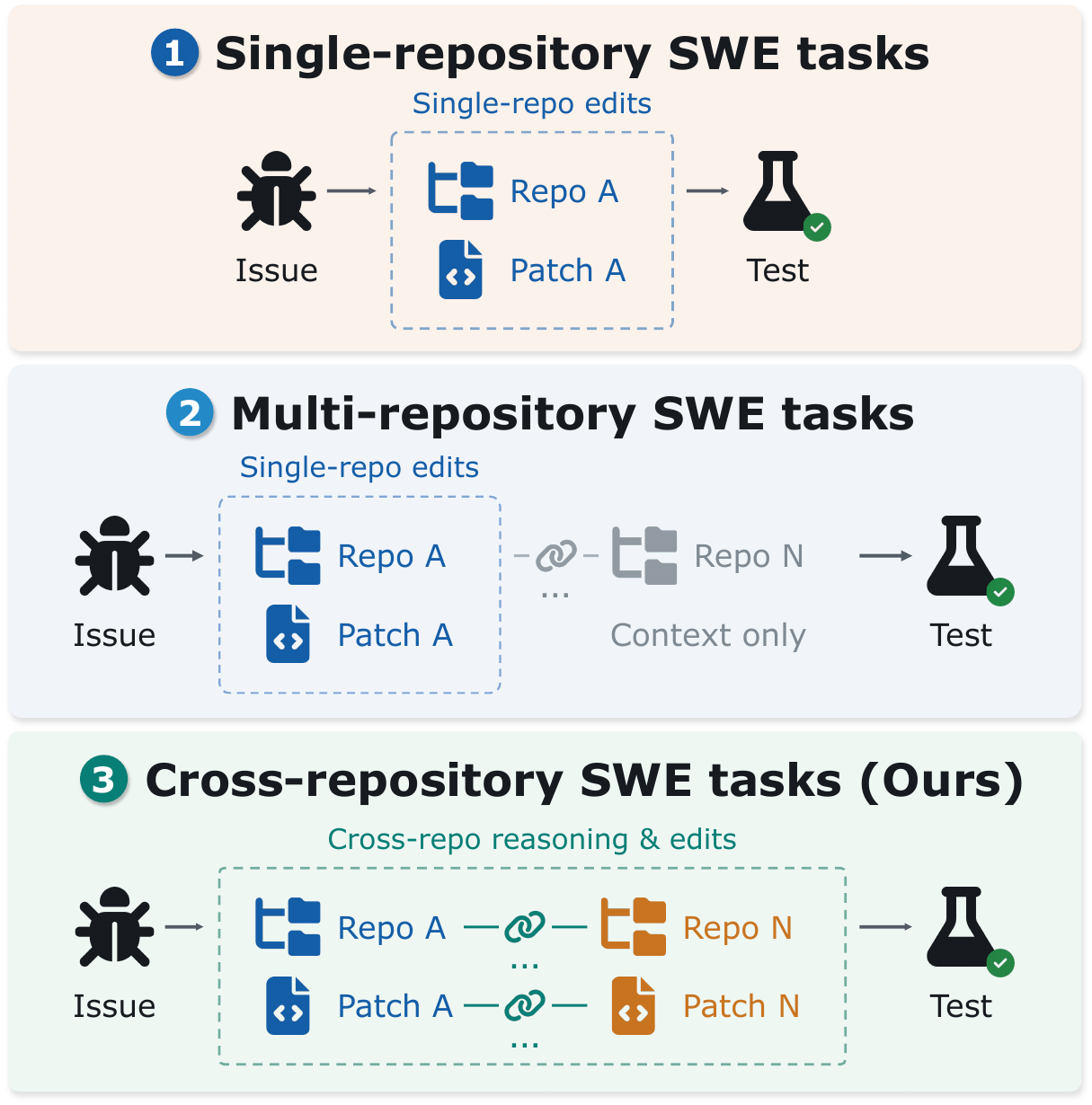}
  \caption{Comparison of single-repository SWE tasks and cross-repository SWE tasks.}
  \label{fig:motivation}
  \vspace{-8pt}
\end{figure}

\begin{figure*}[t]
  \centering
  \includegraphics[width=\textwidth]{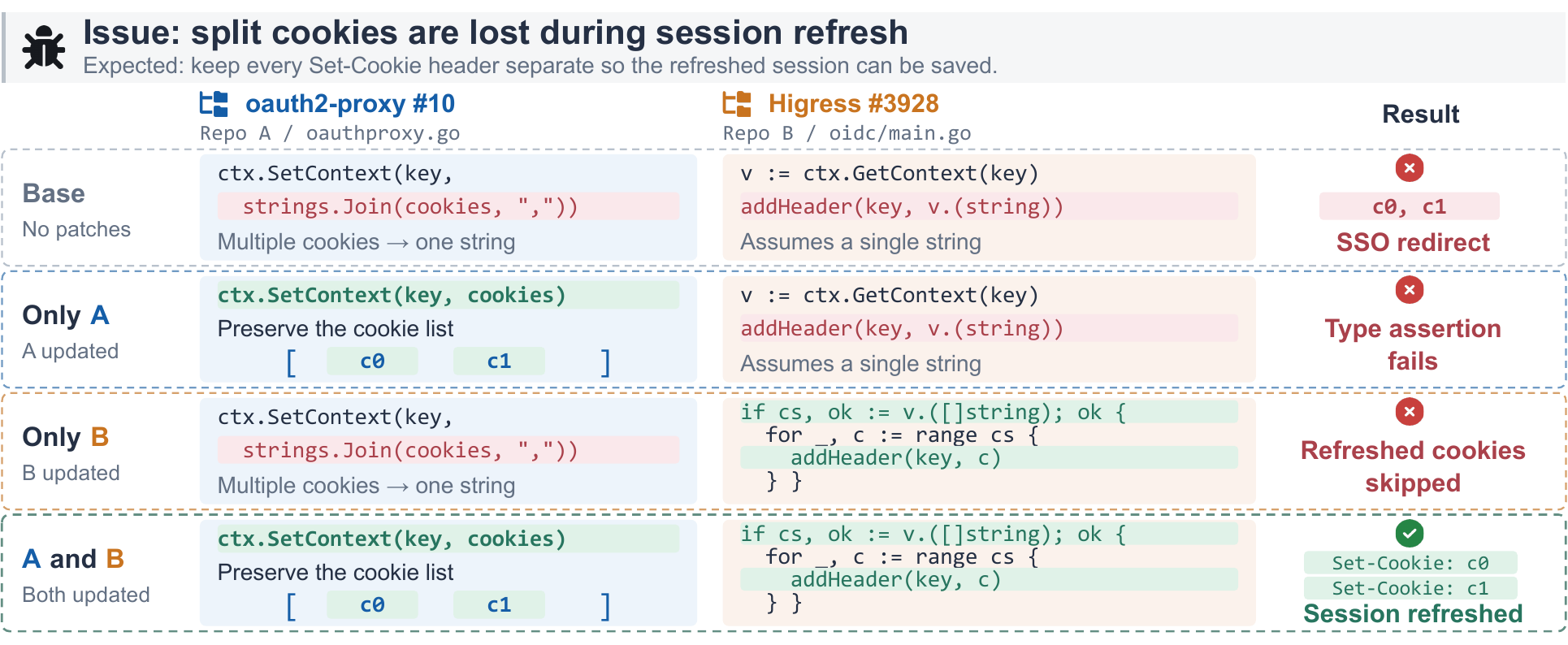}
  \caption{Real-world example of cross-repository joint development involving oauth2-proxy and Higress.}
  \label{fig:case}
  \vspace{-8pt}
\end{figure*}

\section{Introduction}
Large language models (LLMs) and coding agents~\citep{lin2024llmscontinuouslearnersimproving,lin2026knowfixqadrivenrepository,ma2026llmagentscoderepositories,DBLP:journals/corr/abs-2608-18933,DBLP:journals/corr/abs-2608-19854,chen2026schrodingerscoderepositoryllms,pan2026talking,DBLP:journals/corr/abs-2607-10856,DBLP:journals/corr/abs-2602-07900} have made rapid progress on repository-level software engineering. Benchmarks such as \textsc{SWE-bench}~\citep{jimenez2024swebenchlanguagemodelsresolve} have established issue resolution in real-world repositories as a standard evaluation setting, while subsequent work has expanded this paradigm toward broader language coverage~\citep{DBLP:conf/nips/ZanHLCXZLLCZLXC25}, larger codebases~\citep{DBLP:journals/corr/abs-2606-07297}, longer-horizon tasks~\citep{deng2025swebenchproaiagents}, and more complex software changes~\citep{shi2026swe}. As coding agents achieve increasingly high success rates on established issue-resolution benchmarks, increasing task complexity has become an important direction for evaluating their capabilities. 

However, existing efforts primarily pursue this direction within individual repositories.
No matter how large the codebase or how long the task, the context an agent needs to understand and the scope of changes it needs to consider are still bounded by a single repository, which limits how much further such tasks can grow in complexity.
Some recent benchmarks, such as \textsc{BeyondSWE}~\citep{DBLP:journals/corr/abs-2603-03194}  and \textsc{DepBench}~\citep{DBLP:journals/corr/abs-2607-17957}, have started to bring multiple repositories into the picture.
However, the other repositories in these settings merely serve as context.
The agent reads them to understand an API, a dependency, or a change history, but the patch it writes and the tests that check it still stay within one repository.

To introduce greater complexity, we propose cross-repository SWE tasks for agent evaluation, which requires an agent to change multiple repositories together to complete a single task (Figure~\ref{fig:motivation}).
It is a common real-world practice, as changes to APIs, libraries, or protocols in one repository can propagate to other dependent repositories, requiring corresponding modifications across repository boundaries. Such dependencies arise in API evolution, dependency migration~\citep{DBLP:journals/corr/abs-2607-17957}, compatibility fixes, and feature rollouts, where modifying only one repository may be insufficient to achieve the intended behavior. 
Figure~\ref{fig:case} illustrates a real-world example involving two repositories: \texttt{oauth2-proxy} and \texttt{Higress}. During session refresh, multiple cookies are incorrectly merged into a single HTTP header, causing authentication failures. Resolving this issue requires changes to both repositories: \texttt{oauth2-proxy} must preserve cookies as a list, while \texttt{Higress} must process and emit each cookie separately. Modifying only one repository leaves the shared interface inconsistent or the cookies improperly handled. Only the coordinated changes restore the intended behavior, illustrating the cross-repository joint necessity.


\begin{figure*}[t]
  \centering
  \includegraphics[
    width=\textwidth]{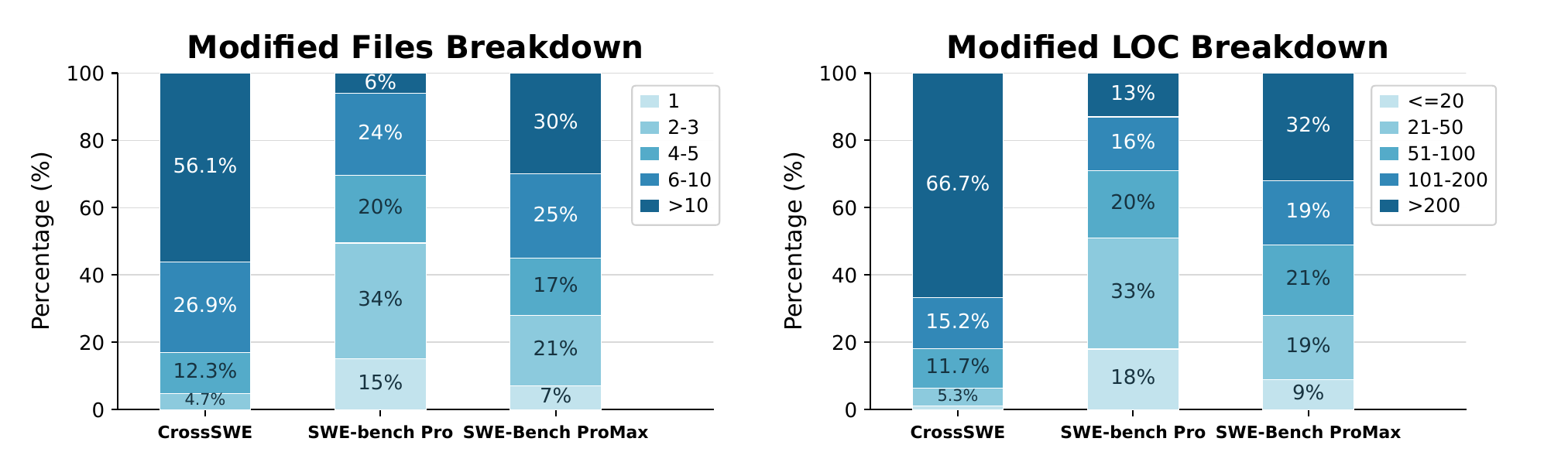}
  \setlength{\abovecaptionskip}{2pt}
  \caption{\small Distribution of reference-patch sizes across \approach{}, SWE-bench Pro, and SWE-Bench ProMax.}
  \label{fig:patch-complexity}
\end{figure*}

Unlike conventional repository-level repair, cross-repository joint development requires agents to reason about dependencies and implementation constraints across multiple codebases~\citep{DBLP:journals/corr/abs-2603-03194}. Beyond understanding the implementation within each repository, agents must identify the relevant cross-repository interfaces, determine how changes affect dependent components, and produce mutually compatible modifications. Even when individual repository patches appear correct, inconsistencies between them can prevent the overall task from being resolved. These requirements introduce additional challenges in dependency reasoning, change propagation, and cross-repository consistency that cannot be fully captured by single-repository evaluation.

To address this gap, we introduce \approach{}, a benchmark for cross-repository joint development comprising 171 real-world instances 
spanning 165 distinct repositories, 20 technical domains, and 13 programming languages. Among these instances, 165 involve two repositories, five involve three repositories, and one involves five repositories. As shown in Figure~\ref{fig:patch-complexity}, \approach{} exhibits a broader reference-patch scope than SWE-bench Pro and SWE-Bench ProMax, with 56.1\% of instances modifying more than ten files and 66.7\% changing over 200 lines of code.

Our contributions are summarized as follows:
(1) We introduce \approach{}, a benchmark comprising 171 real-world cross-repository joint-development tasks spanning 165 repositories, 20 technical domains, and 13 programming languages. (2) We evaluate five frontier LLMs, revealing substantial challenges in cross-repository joint development, with the strongest model achieving only 58.48\% Pass@1. (3) We identify cross-repository inconsistencies as the most prevalent failure mode, highlighting the difficulty of maintaining consistency across interdependent repositories.





\section{Related Work}

\subsection{Repository-Level Software Engineering Benchmarks}

Recent software engineering benchmarks have progressively expanded coding agent evaluation~\citep{chen2026sweatscale,zeng2025pruning,wang2026rsihub,zeng2026dockerless,zeng2026glimprouter,zhang2026paratempo,DBLP:journals/corr/abs-2602-07900,chen2026sweexpexperiencedrivensoftwareissue,shi2026earlyevalcheaperagentevaluation,sun2026executinggeneratehidingexecution} from repository-level issue resolution to broader languages, task scales, and development stages. \textsc{SWE-Bench}~\citep{jimenez2024swebenchlanguagemodelsresolve} established repository-level issue resolution as a realistic benchmark for coding agents, requiring agents to modify real GitHub repositories to complete tasks and pass corresponding test suites. 
\textsc{SWE-bench Pro}~\citep{deng2025swebenchproaiagents} targets more challenging and longer-horizon software engineering tasks, emphasizing realistic development scenarios that often require substantial reasoning and modifications across multiple files. \textsc{SWE-Bench ProMax}~\citep{shi2026swe} further expands the scope toward expert-curated, large-scale software engineering tasks across multiple programming languages, including complex refactoring and code transformation. 
\textsc{FEA-Bench}~\citep{li2025fea} and \textsc{FeatureBench}~\citep{DBLP:conf/iclr/ZhouZWHWHYWPFTZ26} extend repository-level evaluation beyond issue fixing to feature implementation, with the former constructing tasks from real-world pull requests and the latter targeting complex feature development spanning multiple commits and pull requests.

However, these efforts primarily increase complexity within individual repositories, leaving cross-repository coordination largely unexplored. \approach{} explores this complementary dimension by benchmarking real-world development tasks that require jointly necessary changes across multiple interdependent repositories.

\begin{figure*}[t]
  \centering
  \includegraphics[
    width=\textwidth]{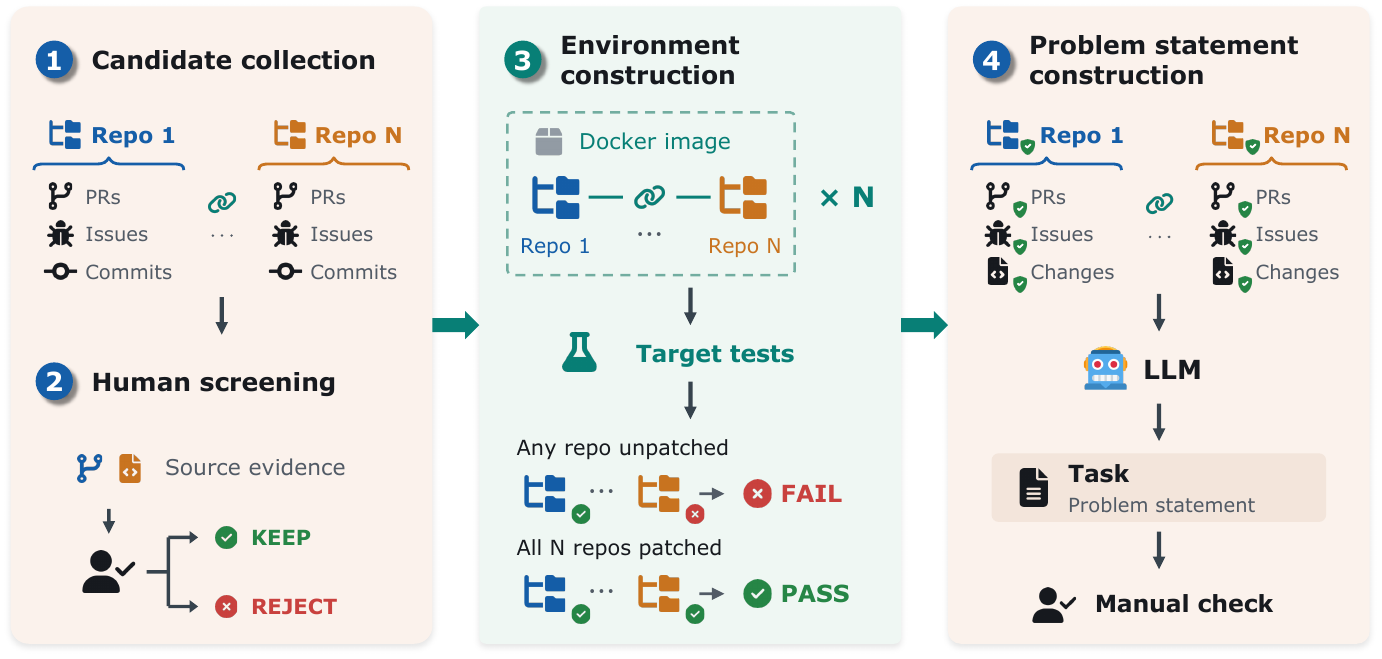}
  \caption{Overview of the \approach{} benchmark construction framework.}
  \label{fig:framework}
\end{figure*}

\subsection{Multi-Repository Software Engineering}

Recent benchmarks~\citep{202605.2065,DBLP:journals/corr/abs-2603-03194,DBLP:journals/corr/abs-2607-17957} have begun to move beyond single-repository issue resolution by considering software-engineering tasks that involve multiple repositories or external development contexts. \textsc{BeyondSWE}~\citep{DBLP:journals/corr/abs-2603-03194} broadens coding-agent evaluation beyond single-repository bug fixing along both resolution scope and knowledge scope, covering settings such as cross-repository reasoning, dependency-driven migration, domain-specialized problem solving, and full-repository generation. 
\textsc{DepBench}~\citep{DBLP:journals/corr/abs-2607-17957} further studies dependency breaking changes as a cross-repository repair problem, where upstream release notes and API changes provide evidence for repairing a downstream consumer after a dependency update. However, the upstream repository primarily serves as a source of repair evidence, while the generated patch and executable validation remain confined to one repository.

Although these efforts extend the scope of repository-level evaluation, incorporating cross-repository information or dependencies does not necessarily require multiple repositories to be modified together. In contrast, \approach{} introduces \emph{cross-repository joint development} as a complementary dimension of task complexity, where the solution itself spans multiple interdependent repositories. By explicitly verifying the joint necessity of repository changes, \approach{} evaluates whether coding agents can implement mutually compatible modifications across repository boundaries rather than merely reason about cross-repository dependencies.

\section{\approach}
This section describes how the benchmark is constructed, verified, and composed.
\subsection{Task Formulation}

Each instance in \approach{} consists of four components: 
(1) a pre-configured execution environment containing two or more participating repositories at their respective base commits, with all required dependencies installed; 
(2) a problem statement describing a single software-engineering task that spans these repositories; 
(3) a test suite that validates the target cross-repository behavior as well as required existing behavior; and 
(4) repository-specific gold patches recording the original developers' coordinated solution. 
Given the environment and the problem statement, an agent must autonomously inspect and modify the participating repositories so that all required tests pass. 
An instance is considered \textit{resolved} if and only if the resulting joint repository state satisfies the complete evaluation suite. 
Unlike conventional repository-level tasks, the solution in \approach{} is not confined to a single codebase: the intended behavior requires coordinated changes across multiple repositories, such that modifying any repository in isolation is insufficient. 

\subsection{Benchmark Collection}


Figure~\ref{fig:framework} illustrates the construction pipeline of \approach{}, which progresses from broad candidate discovery to source-grounded validation and reproducible environment construction. 
We adopt a multi-stage filtering and validation pipeline that progressively improves source validity, reproducibility, and task quality.

\begin{figure*}[t]
  \centering
  \includegraphics[width=\textwidth]{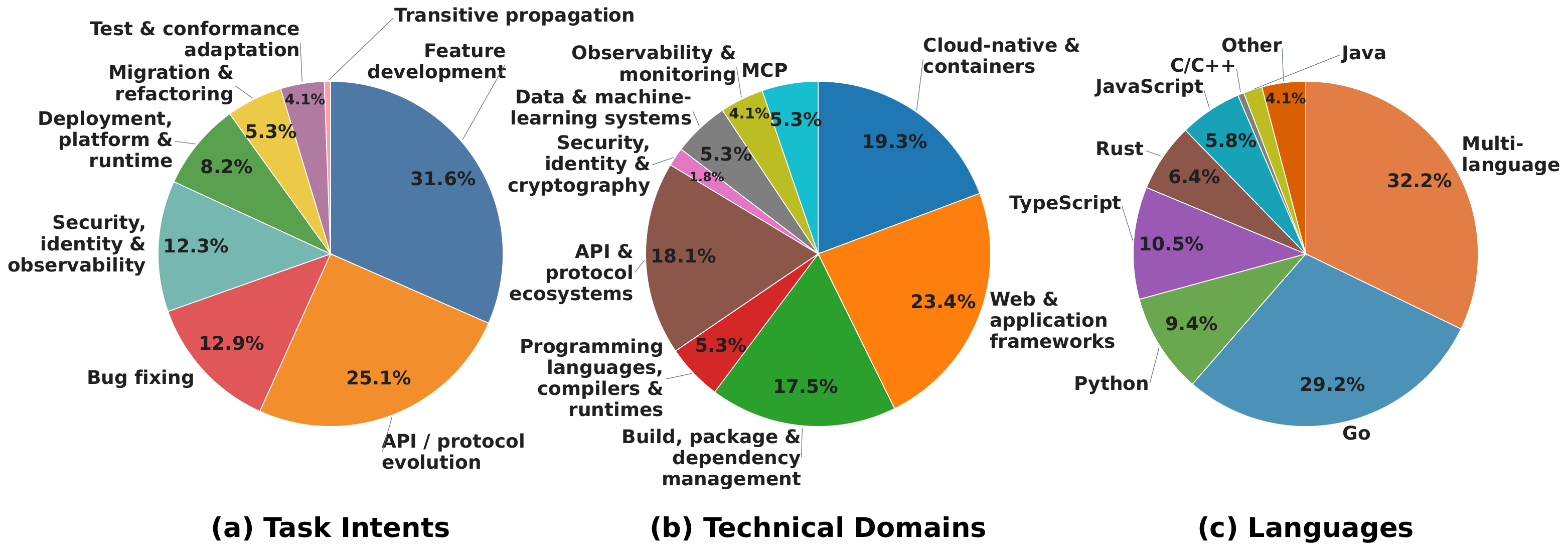}
  \caption{Diversity of \approach{} in terms of task intents, technical domains, and programming languages.}
  \label{fig:benchmark-diversity}
\end{figure*}

\textbf{Stage 1: Candidate collection.}
We use the GitHub API to identify active repositories and search their development histories for candidate cross-repository changes from January 2025 through September 2026. 
We mine merged pull requests, issues with reconstructable changes, commits, and release or compatibility discussions using cross-repository relationship cues such as API evolution, dependency migration, provider--consumer, and client--server relations. We further identify candidates through explicit references to issues, pull requests, commits, or repositories in other codebases, which serve as signals of potential cross-repository coordination. 

\textbf{Stage 2: Human screening.}
Since the discovery stage is intentionally designed for high recall, we next manually screen the collected candidates to remove cases that do not represent genuine cross-repository development. We inspect each candidate together with its source pull request or issue, development timeline, referenced repositories, and corresponding commits. A candidate is retained only when the source evidence identifies a concrete software change and at least two repositories contain substantive implementation changes that are causally, dependently, or coordinately related to the same task. We exclude candidates in which the additional repository involvement is limited to documentation, release notes, changelogs, or other non-implementation changes. Candidates with insufficient or ambiguous evidence are discarded. The remaining candidates proceed to multi-repository environment construction.

\textbf{Stage 3: Environment construction and joint-necessity verification.}
For each retained candidate, we construct a dedicated Docker image containing all repositories involved in the task. Each repository is checked out at its verified base commit, and all dependencies required to build and execute the repositories are installed in the same environment. We require every checkout to remain in a clean Git state and exclude fixed checkouts, gold patches, hidden tests, and artifacts produced from the fixed versions from the agent-facing image. This gives each instance a single, reproducible execution environment from which all subsequent evaluations are performed.

We then reconstruct a repository-specific gold patch for each participating repository and use these patches to verify whether the task genuinely requires coordinated cross-repository changes. For an instance involving $N$ repositories, each repository has two possible states: its original base state or the state obtained after applying its gold patch. We exhaustively enumerate the resulting $2^N$ mixed-version configurations within the same Docker environment, changing only which repository-specific patches are applied. For example, an instance involving two repositories $A$ and $B$ produces four configurations: $(A_0,B_0)$, $(A_1,B_0)$, $(A_0,B_1)$, and $(A_1,B_1)$, where $0$ and $1$ denote the base and patched states, respectively.
We execute the target FAIL\_TO\_PASS tests in every configuration. A candidate satisfies our joint-necessity criterion only when the target behavior fails in the all-base configuration and in every partially patched configuration, but succeeds when all required repository patches are applied together. Thus, for the two-repository case, $(A_0,B_0)$, $(A_1,B_0)$, and $(A_0,B_1)$ must fail, while $(A_1,B_1)$ must pass. This explicitly distinguishes tasks whose solution merely involves multiple repositories from tasks for which the repository changes are jointly necessary. 

\begin{table*}[t]
\centering
\small

\begin{tabular}{lccccccc}
\toprule
\multirow{2}{*}{Model}
& \multicolumn{3}{c}{\textbf{Pass@1 (\%) $\uparrow$}}
& \multirow{2}{*}{\makecell{Avg. Input\\Tokens}}
& \multirow{2}{*}{\makecell{Avg. Output\\Tokens}}
& \multirow{2}{*}{Avg. Steps}
& \multirow{2}{*}{Avg. Cost (\$)} \\
\cmidrule(lr){2-4}

& Overall
& Single-lang.
& Multi-lang.
& & & & \\

\midrule

\rowcolor{gray!20}
\multicolumn{8}{c}{\textit{Closed-source Models}} \\

GPT-6.1 Sol
& \textbf{58.48}
& \textbf{60.34}
& \textbf{54.55}
& 0.97M
& 12.5K
& 32.9
& 0.3496 \\

Claude Sonnet 5.5
& 50.29
& 52.59
& 43.64
& 1.31M
& 16.3K
& \textbf{29.2}
& 0.5217 \\

\midrule

\rowcolor{gray!20}
\multicolumn{8}{c}{\textit{Open-source Models}} \\

DeepSeek V4.1 Flash
& 52.63
& 52.59
& 52.73
& 17.54M
& 88.73K
& 152.7
& \textbf{0.2881} \\

GLM-5.3 Flash
& 43.86
& 45.69
& 40.00
& 9.74M
& 55.21K
& 141.7
& 0.3411 \\

Qwen3.8-Flash-Next
& 45.03
& 45.61
& 43.86
& 13.31M
& 75.68K
& 182.1
& 0.2939 \\

\bottomrule
\end{tabular}

\caption{Overall performance of different models on \approach{}.}
\label{tab:overall-performance}
\end{table*}

\textbf{Stage 4: Problem statement construction.}
The development artifacts associated with a candidate---such as pull requests, issue discussions, commit messages, and release notes---are often fragmented and implementation-oriented. Following SWE-Bench ProMax~\citep{shi2026swe}, we transform each verified candidate into a self-contained natural-language task specification. Drawing on source issues, pull requests, and repository changes, we describe the observable failure, expected behavior, and cross-repository requirements from the pre-fix perspective without prescribing implementation details. We further remove solution-specific information, including gold patches, fixed commits, and evaluator metadata, and manually verify consistency among the problem statement, source evidence, and tests. 
In particular, we explicitly specify the required cross-repository behavior to avoid ambiguity about the intended task scope and ensure that hidden tests evaluate observable behavior rather than gold-patch-specific implementation details, such as function names, field names, or file paths.

\subsection{Benchmark Composition}

\approach{} contains 171 verified instances spanning diverse cross-repository development scenarios. We characterize the benchmark along three dimensions: task intents, technical domains, and programming-language diversity. 

\textbf{Task complexity.}
Figure~\ref{fig:patch-complexity} compares the reference-patch sizes of \approach{} with SWE-bench Pro and SWE-Bench ProMax. In \approach{}, 56.1\% of instances modify more than ten files, compared with 6\% in SWE-bench Pro and 30\% in SWE-Bench ProMax. Similarly, 66.7\% of instances involve more than 200 modified lines, compared with 13\% and 32\%, respectively. These results highlight the broader aggregate solution scope of \approach{}, where coding agents must implement changes across multiple repositories rather than within a single repository.

\textbf{Task and ecosystem diversity.}
Beyond the scale of required modifications, we further examine the diversity of software-engineering scenarios covered by \approach{}.
Figure~\ref{fig:benchmark-diversity} illustrates the diversity of \approach{} across task intents, technical domains, and programming languages. The benchmark covers eight task-intent categories, including feature development, API/schema/protocol evolution, security and observability coordination, migration, and compatibility maintenance. These tasks span 20 technical domains and 13 programming languages, covering heterogeneous software ecosystems and cross-language development scenarios. This diversity enables \approach{} to assess coding agents across different development objectives, technical stacks, and language boundaries, providing a broader evaluation of cross-repository coordination capabilities.

\section{Experimental Setup}

\paragraph{Implementation Details.}
We use \textsc{mini-SWE-agent} v2.4.6~\cite{DBLP:conf/nips/YangJWLYNP24} as the common agent scaffold for all experiments. \textsc{mini-SWE-agent} is a lightweight repository-level coding-agent framework that enables language models to inspect, modify, and test codebases through iterative shell-based interactions. We evaluate five representative frontier models: GPT-6.1 Sol~\cite{openai2026gpt61sol}, Claude Sonnet 5.5~\cite{anthropic2026sonnet55}, DeepSeek V4.1 Flash~\cite{deepseekai2026deepseekv41flash}, GLM-5.3 Flash~\cite{DBLP:journals/corr/abs-2602-15763} and Qwen3.8-Flash-Next~\cite{qwen3.8flashnext}. For all models, we use the \textit{high} reasoning-effort setting and keep the agent configuration fixed across experiments. Each run is limited to at most 300 interaction steps and a maximum model cost of \$10.

\paragraph{Evaluation Metrics.}
We evaluate each model on the same set of \approach{} instances and report \textbf{Pass@1} as the primary effectiveness metric, where an instance is counted as resolved only when the final repository state satisfies the complete evaluation suite. We additionally report \textbf{Avg. Input Tokens}, \textbf{Avg. Output Tokens}, \textbf{Avg. Steps}, and \textbf{Avg. Cost (\$)} to characterize the computational and interaction cost of each agent.

\begin{figure*}[t]
  \centering
  \includegraphics[width=\textwidth]{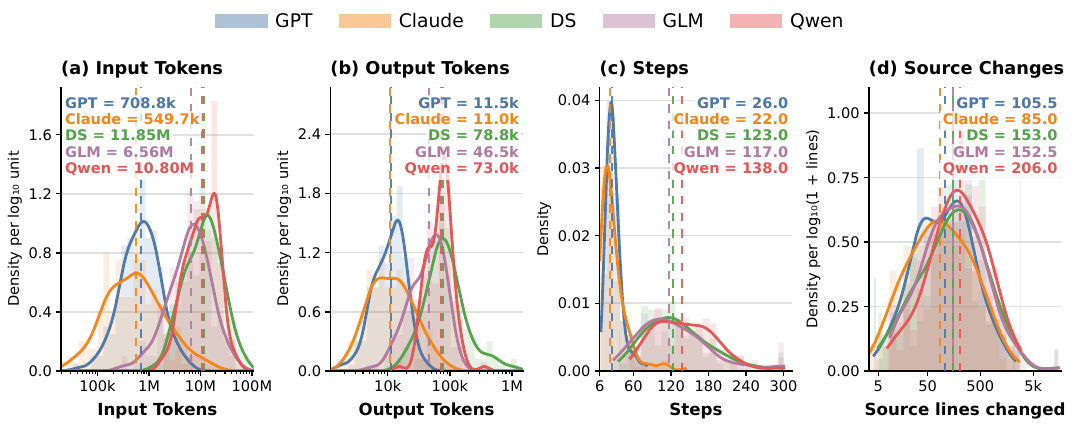}
    \caption{Distributions of token usage, code changes, and interaction steps across five coding agents on \approach{}. Dashed lines indicate model-wise medians.}
  \label{fig:distributions}
\end{figure*}

\section{Results}

\subsection{Main Results}

\textbf{\approach{} remains challenging even for the strongest evaluated coding agents.}
Table~\ref{tab:overall-performance} summarizes the overall performance and resource usage of the evaluated coding agents on \approach{}. Even the strongest model, GPT-6.1 Sol, achieves only 58.48\% Pass@1 on \approach{}, substantially below its reported 75.2\% performance on DeepSWE v1.1. This gap suggests that cross-repository joint development introduces substantial challenges beyond conventional repository-level software engineering tasks. Among the remaining models, DeepSeek V4.1 Flash achieves 52.63\%, followed by Claude Sonnet 5.5 (50.29\%), GLM-5.3 Flash (43.86\%) and Qwen3.8-Flash-Next (45.03\%). Performance also varies across language settings. GPT-6.1 Sol leads on single-language tasks with 60.34\% Pass@1, compared with 54.55\% on multi-language tasks. Notably, DeepSeek V4.1 Flash remains competitive on multi-language tasks, achieving 52.73\% Pass@1. These results reveal distinct model capabilities when cross-repository development spans programming-language boundaries.

\textbf{Closed-source models achieve competitive costs despite higher token prices.}
Despite its higher per-token pricing, GPT-6.1 Sol achieves the highest Pass@1 with an average cost of \$0.3496 per instance, close to GLM-5.3 Flash (\$0.3411) and moderately higher than DeepSeek V4.1 Flash (\$0.2881) and Qwen3.8-Flash-Next (\$0.2939). This cost competitiveness is consistent with its substantially shorter interaction trajectories. GPT-6.1 Sol and Claude Sonnet 5.5 require only 32.9 and 29.2 steps on average, respectively, compared with 152.7 for DeepSeek V4.1 Flash, 141.7 for GLM-5.3 Flash, and 182.1 for Qwen3.8-Flash-Next. These results highlight GPT-6.1 Sol's favorable effectiveness--cost trade-off, achieving the highest task success rate while offsetting its higher per-token pricing through substantially shorter interaction trajectories.

\subsection{Joint vs. Independent Development}

To investigate the additional challenges introduced by joint cross-repository development, we derive 330 single-side modification tasks from the 165 two-repository instances in \approach{}. Each derived task retains both repositories in the execution environment but requires modifications to only one repository, allowing us to examine agent performance when cross-repository context is preserved but joint modification is no longer required.

\paragraph{Joint cross-repository development is more challenging than single-side modification.}
As shown in Table~\ref{tab:joint-vs-single}, both models achieve higher Pass@1 with fewer actions under the single-side setting. GPT-6.1 Sol improves from 58.48\% to 64.55\% (+6.07 percentage points), while its average steps decrease from 32.90 to 18.54. Similarly, DeepSeek V4.1 Flash improves from 52.63\% to 61.81\% (+9.18\%), with average steps decreasing from 152.70 to 126.94. These results indicate that cross-repository joint development poses additional challenges beyond accessing multiple repositories, as agents must implement mutually compatible changes across repository boundaries.

\begin{figure*}[t]
  \centering
  \includegraphics[width=0.92\textwidth]{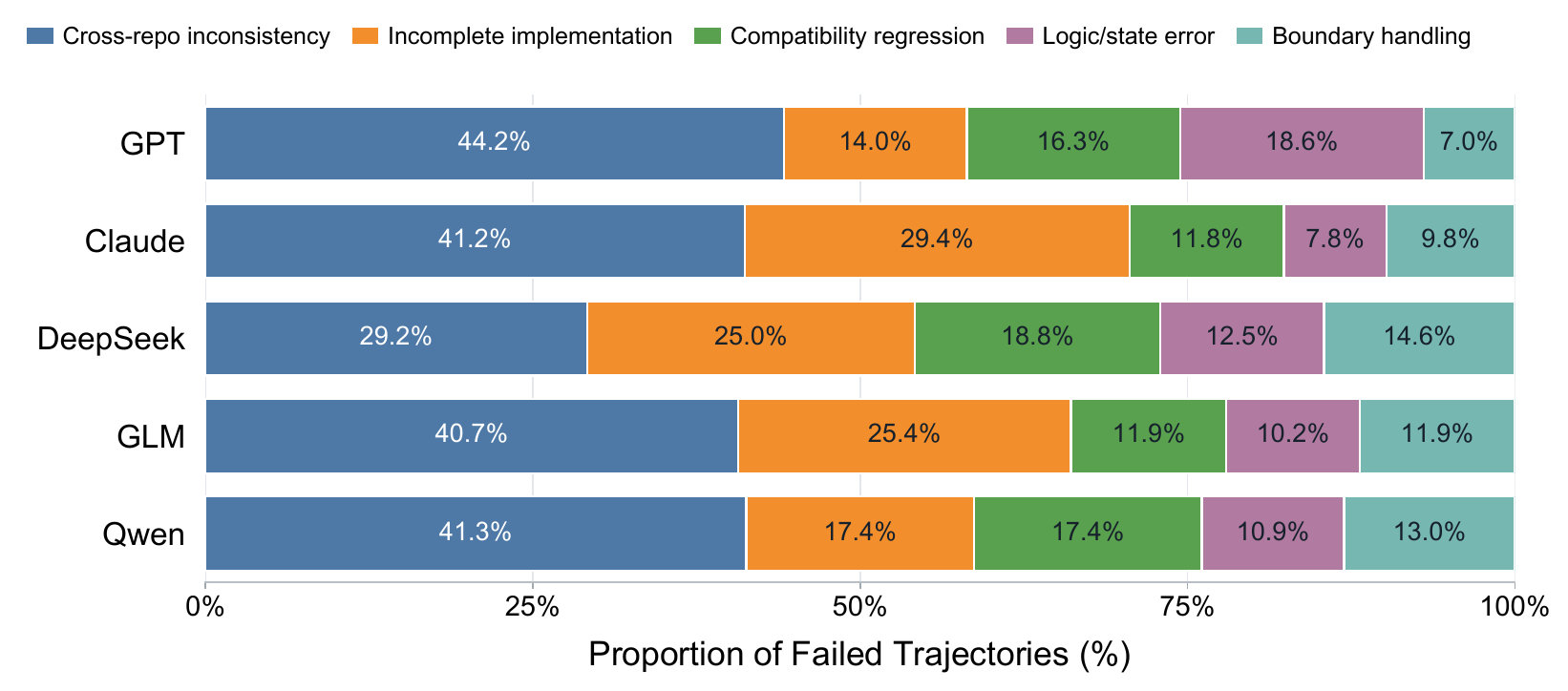}
  \caption{Distribution of primary failure modes across five coding agents on \approach{}.}
  \label{fig:failure-modes}
\end{figure*}

\begin{table}[t]
\centering
\scriptsize
\setlength{\tabcolsep}{3pt}
\renewcommand{\arraystretch}{1.18}

\resizebox{\columnwidth}{!}{
\begin{tabular}{lcccc}
\toprule
\multirow{2}{*}{Model}
& \multicolumn{2}{c}{Pass@1 (\%) $\uparrow$}
& \multicolumn{2}{c}{Avg. steps $\downarrow$} \\
\cmidrule(lr){2-3}
\cmidrule(lr){4-5}
& Joint & Single-side
& Joint & Single-side \\
\midrule

GPT-6.1 Sol
& 58.48
& \textbf{64.55} \textcolor{green!50!black}{(+6.07)}
& 32.90
& \textbf{18.54} \textcolor{green!50!black}{(-14.36)} \\

DeepSeek V4.1 Flash
& 52.63
& \textbf{61.81} \textcolor{green!50!black}{(+9.18)}
& 152.70
& \textbf{126.94} \textcolor{green!50!black}{(-25.76)} \\

\bottomrule
\end{tabular}
}

\caption{Comparison of joint and single-side repository development on \approach{}.}
\label{tab:joint-vs-single}
\end{table}

\subsection{Agent Behaviour Analysis}

Figure~\ref{fig:distributions} compares the distributions of input tokens, output tokens, final source lines changed, and interaction steps across the five coding agents on \approach{}. We further examine their trajectories to understand how different agents approach cross-repository development tasks.

\textbf{Fewer interaction steps do not imply proportionally smaller code modifications.}
Although GPT-6.1 Sol and Claude Sonnet 5.5 require substantially fewer steps, the differences in final source lines changed are much smaller than those in interaction length. This suggests that their shorter trajectories are not simply the result of making fewer modifications, but may reflect a greater concentration of code changes within fewer interactions. In contrast, DeepSeek V4.1 Flash and GLM-5.3 Flash perform substantially more interactions without a proportional increase in the size of their final patches.

\textbf{Coding agents exhibit distinct exploration and modification strategies.}
Our manual inspection of agent trajectories reveals different problem-solving patterns. GPT-6.1 Sol and Claude Sonnet 5.5 tend to identify relevant code locations and begin modifying repository source code earlier, often making larger, more concentrated edits. In contrast, DeepSeek V4.1 Flash and GLM-5.3 Flash tend to spend more interactions exploring repository structures and dependencies, followed by frequent alternation between incremental code modifications and testing. 
Such differences may help explain why longer interaction trajectories do not consistently yield better cross-repository task performance.

\subsection{Failure Patterns Analysis}

To investigate why coding agents fail on cross-repository joint-development tasks, we manually analyze unsuccessful trajectories and classify their primary failure causes into five categories: (1) \textit{Cross-repository inconsistency}, involving incompatible interfaces, protocols, or artifacts across repositories, including related compilation and build failures; (2) \textit{Incomplete implementation}, where the agent fails to implement all required functionality; (3) \textit{Compatibility regression}, where modifications break previously correct behavior; (4) \textit{Logic/state errors}, involving incorrect program logic or state management; and (5) \textit{Boundary handling}, involving unhandled edge cases or exceptional inputs.

\paragraph{Cross-repository inconsistencies dominate agent failures across all five models.}
As shown in Figure~\ref{fig:failure-modes}, cross-repository inconsistency is the most prevalent failure category across all five models, accounting for 44.2\% of GPT's unsuccessful trajectories, 41.2\% of Claude's, 29.2\% of DeepSeek's, and 40.7\% of GLM's. Incomplete implementation is another major failure mode, particularly for Claude (29.4\%), DeepSeek (25.0\%), and GLM (25.4\%). Meanwhile, DeepSeek exhibits a more distributed failure profile, with relatively higher proportions of compatibility regressions (18.8\%) and boundary-handling errors (14.6\%). These findings suggest that, although models differ in their failure patterns, maintaining consistency across repositories remains a common challenge for current coding agents.


\section{Conclusion}
In this paper, we introduce \approach{}, a benchmark that extends software-engineering evaluation beyond single-repository task complexity to cross-repository joint development. It comprises 171 real-world instances spanning 165 repositories, 20 technical domains, and 13 programming languages, with executable verification of jointly necessary repository changes. Our evaluation of five LLMs reveals substantial challenges, with the strongest model achieving only 58.48\% Pass@1. 
Comparisons with single-side modification tasks further reveal higher success rates when only one repository requires changes, highlighting the additional challenges of joint development. 
Failure analysis further identifies cross-repository inconsistencies as the dominant failure mode, highlighting the limitations of current coding agents in maintaining consistency across repository boundaries. \approach{} provides a foundation for evaluating and improving cross-repository development capabilities.

\newpage
\section*{Limitations}

Despite its contributions, \approach{} has several limitations. First, although the benchmark contains 171 real-world instances spanning 165 repositories and 20 technical domains, its coverage of large-scale multi-repository development remains limited. Specifically, 165 instances involve two repositories, while only six involve three or more repositories. Future work could extend the benchmark to larger networks of interdependent repositories.

Second, \approach{} provides agents with the participating repositories to focus evaluation on cross-repository dependency reasoning and coordinated code modifications rather than repository discovery. However, real-world development may require agents to identify relevant repositories from a broader software ecosystem. Future extensions could introduce irrelevant repositories to assess repository discovery and selection capabilities.

Third, we adopt a unified mini-SWE-agent scaffold and consistent configurations to ensure fair comparisons across models. Since mini-SWE-agent primarily relies on Bash-based interactions, which are widely supported by coding-agent harnesses, our findings may also provide insights into other shell-based frameworks, although direct validation remains necessary. We additionally impose limits of 300 interaction steps and \$10 per instance.

\section*{Ethical Considerations}

Our benchmark is constructed from publicly available GitHub repositories and development histories. Potential risks include the redistribution of personal information, sensitive data, or offensive content contained in source artifacts, as well as the computational resources required for agent evaluation. To mitigate these risks, we reviewed the benchmark artifacts before release to identify personal identifiers, sensitive information, and potentially offensive or objectionable content. We focus on technical information necessary for task reconstruction and evaluation rather than developers' personal identities or unrelated content, minimizing unnecessary exposure of such information. Additionally, evaluating coding agents may incur substantial computational and monetary costs, which should be considered when reproducing or extending our experiments.

\bibliography{ref}

\newpage
\appendix
\section{Appendix}

\subsection{Data Availability}
We publicly release \approach{} to support future research on cross-repository software engineering: \url{https://huggingface.co/datasets/CrossSWE/CrossSWE-Bench}.

\subsection{Task-Intent Taxonomy}
The eight high-level task-intent categories are mutually exclusive and cover all 171 instances in the formal evaluation subset. The definitions shown in table~\ref{tab:intent-taxonomy} describe the primary cross-repository behavior represented by each high-level category.

\begin{table*}[t]
    \centering
    \small
    \setlength{\tabcolsep}{5pt}
    \renewcommand{\arraystretch}{1.08}
    \caption{High-level task-intent taxonomy used for the formal 171-instance benchmark composition analysis.}
    \label{tab:intent-taxonomy}
    \begin{tabular}{p{0.25\textwidth} r p{0.60\textwidth}}
        \hline
        \textbf{Category} & \textbf{Instances} & \textbf{Definition} \\
        \hline
        Feature development & 54 & Implement a new user-facing capability whose coordinated behavior spans multiple repositories. \\
        API / schema / protocol evolution & 43 & Change a shared API, schema, wire protocol, or interface contract together with its clients, servers, or consumers. \\
        Bug fixing & 22 & Repair incorrect cross-repository behavior, build or release configuration, or downstream compatibility while preserving required behavior. \\
        Security, identity \& observability coordination & 21 & Coordinate authentication, authorization, identity, telemetry, logging, or monitoring changes across repositories. \\
        Deployment, platform \& runtime coordination & 14 & Align deployment configuration, platform integration, or runtime behavior across repositories. \\
        Migration, upgrade \& refactoring & 9 & Perform a synchronized dependency, version, architecture, or code-organization transition. \\
        Test, benchmark \& conformance adaptation & 7 & Coordinate tests, benchmarks, or conformance assets with the implementation changes they validate. \\
        Transitive change propagation & 1 & Propagate a required change through a multi-step dependency chain involving more than two repositories. \\
        \hline
    \end{tabular}
\end{table*}

\section{Data Licensing and Release}

\approach{} is constructed from publicly available GitHub repositories and development histories. The benchmark dataset is hosted on Hugging Face, with source references and patch provenance preserved for attribution. We reviewed the upstream licenses of all 165 repositories and documented their applicable terms in the dataset release. Original non-code benchmark materials are licensed under CC-BY-4.0, while third-party source-code contributions remain subject to their respective upstream licenses. For instances involving commercially licensed MUI X components, official licensing documentation permits qualifying non-production benchmark use.

\section{Use of AI Assistants}
AI assistants (such as ChatGPT) were used to draft basic code and refine the language of the manuscript. All experimental designs, data analyses, and conclusions were conducted and verified by the authors themselves. The role of AI was limited to generating general code and optimizing language; it did not influence the research results or their reproducibility.

\end{document}